\documentclass{elsart}
\usepackage{amssymb}

\begin{document} 

\journal{Physics Letters A}

\begin{frontmatter}

\title{Periodic orbit quantization of chaotic maps
by harmonic inversion}
\author{Kirsten Weibert, J\"org Main, and G\"unter Wunner}
\address{Institut f\"ur Theoretische Physik 1,
         Universit\"at Stuttgart, D-70550 Stuttgart, Germany}
\maketitle

\begin{abstract}
A method for the semiclassical quantization of chaotic maps is proposed,
which is based on harmonic inversion.
The power of the technique is demonstrated for the baker's map as a
prototype example of a chaotic map.
\end{abstract}

\end{frontmatter}

The harmonic inversion method for signal processing \cite{Wal95,Man97b}
has proven to be a powerful tool for the semiclassical quantization of 
chaotic as well as integrable dynamical systems \cite{Mai97b,Mai99a,Wei00}.
Starting from Gutzwiller's trace formula for chaotic systems, or the 
Berry-Tabor formula for integrable systems \cite{Gut90}, the harmonic 
inversion method is able to circumvent the convergence problems of the 
periodic orbit sums and to directly extract the semiclassical eigenvalues 
from a relatively small number of periodic orbits. 
The technique has successfully been applied to a large variety of
Hamiltonian systems \cite{Mai99a,Wei00}.
It has been shown that the method is universal in the sense that
it does not depend on any special properties of the dynamical system.

In this Letter we demonstrate that the range of application of the harmonic 
inversion method extends beyond Hamiltonian systems also to quantum maps. 
Starting from the analogue of Gutzwiller's trace formula for chaotic maps, 
we show that the semiclassical eigenvalues of chaotic maps can be determined 
by a procedure very similar to the one for flows.
As an example system we consider the well known baker's map.
For this map we can take advantage of the fact that the periodic orbit 
parameters can be determined analytically.

We briefly review the basics of quantum maps that are relevant to what
follows (for a detailed account of quantum maps see, e.g.\ Ref.~\cite{Haa01}).
We consider quantum maps, acting on a finite dimensional Hilbert space of 
dimension $N$, which possess a well-defined classical limit for $N\to\infty$.
The quantum dynamics is determined by the equation
\begin{equation}
 \psi_{n+1} = U\psi_n \; ,
\end{equation}
where $U$ is a unitary matrix of dimension $N$, and $\psi_n$ is the 
$N$-dimensional discretized wave vector.
The eigenvalues $u_k$ of $U$ lie on the unit circle,
$u_k=\exp(-{\rm i}\varphi_k)$. The density of eigenphases $\varphi_k$ 
on the unit circle is given by
\begin{equation}
\label{rho}
 \rho_{\rm qm}({\varphi})= {N\over 2\pi}+{1\over \pi}\, {\rm Re}
 \sum_{n=1}^{\infty}{\rm Tr}\,U^n\ {\rm e}^{{\rm i}n\varphi}\; ,
\end{equation}
which can be rewritten as
\begin{equation}
 \rho_{\rm qm}({\varphi}) = -{1\over\pi}\ {\rm Im}\,g_{\rm qm}(\varphi)
\end{equation}
with the response function $g_{\rm qm}$ given by
\begin{equation}
\label{gqm1}
 g_{\rm qm}(\varphi) = g_0(\varphi)
 -{\rm i}\sum_{n=1}^{\infty}{\rm Tr}\,U^n\ {\rm e}^{{\rm i}n\varphi} \; .
\end{equation}
In analogy with the periodic orbit theory for flows, 
a semiclassical approximation to the response function (\ref{gqm1})
can be obtained in terms of the periodic orbits of the corresponding
classical system.
[In the context of maps, ``periodic orbit'' means a sequence of fixed points
periodic after $n$ iterations; cyclic shifts of the same sequence
correspond to the same periodic orbit.]
In the semiclassical approximation for maps, the traces of $U^n$ are 
related to the periodic orbits of topological length $n$,
\begin{equation}
\label{trU}
 {\rm Tr}\,U^n \approx\sum_{{\rm po}(n)}
 {n_0\over |{\rm det}(M_{\rm po}-1)|^{1/2}}
 \,{\rm e}^{{\rm i}(S_{\rm po}/\hbar-\mu_{\rm po}\pi /2)}
 =: {\rm i}{\mathcal A}_n\; ,
\end{equation} 
where the sum runs over all periodic orbits of topological length $n$ 
including multiple traversals of shorter orbits.
Here, $S_{\rm po}$ is the action associated with the periodic orbit,
$\mu_{\rm po}$ is its Maslov index, $M_{\rm po}$ is the monodromy matrix 
of the orbit, and $n_0$ is the topological length of the underlying 
primitive orbit (i.e., the length of the shortest subperiod).
The value of the Planck constant is related to the dimension of the
Hilbert space via $\hbar= 1/(2\pi N)$.
The accuracy of the semiclassical approximation can therefore
be expected to improve with increasing $N$.

The central idea for applying harmonic inversion to the periodic orbit 
quantization of maps now is to adjust the semiclassical response function
\begin{equation}
\label{gsemi}
 g({\varphi})= g_0(\varphi)
 + \sum_n{\mathcal A}_n\, {\rm e}^{{\rm i}n\varphi}\; 
\end{equation}
to the form of the exact quantum response function (\ref{gqm1}), expressed 
in terms of the eigenphases $\varphi_k$ and their multiplicities $m_k$,
\begin{equation}
\label{gqm}
 g_{\rm qm}(\varphi) = \sum_k{m_k\over \varphi-\varphi_k} \; .
\end{equation}
It should be pointed out that for all maps by virtue of (\ref{gqm1}) the 
semiclassical amplitudes ${\mathcal A}_n$ are independent of the phase 
$\varphi$.
In analogy with the harmonic inversion procedures for Hamiltonian 
flows \cite{Mai97b,Mai99a}, we Fourier transform the oscillating part 
of the semiclassical response function (\ref{gsemi}) to obtain the 
semiclassical signal
\begin{equation}
\label{Csemi}
 C(s) = \sum_n{\mathcal A}_n\, \delta(s-n)\;.
\end{equation}
The eigenphases $\varphi_k$ can now be determined by adjusting the 
semiclassical signal (\ref{Csemi}) to the form of the corresponding exact 
quantum signal (the Fourier transform of the exact response function 
(\ref{gqm}))
\begin{equation}
 C_{\rm qm}(s) = -{\rm i}\sum_k m_k\, {\rm e}^{-{\rm i}s\varphi_k}
\end{equation}
by harmonic inversion.
Note that compared to the corresponding procedure for flows, the topological 
length now plays the r\^ole of the scaled action, while the actions 
$S_{\rm po}$ of the orbits are included in the amplitudes ${\mathcal A}_n$.
Therefore, all orbits of the same topological length $n$ contribute to the 
same signal point.
While for flows the semiclassical signal takes on the simple
form of a sum over $\delta$ functions only if we assume a certain
scaling property \cite{Mai99a,Wei00}, the form of the signal for maps 
is always the same, as the amplitudes ${\mathcal A}_n$ are for all maps 
independent of $\varphi$.

We will now apply the general procedure discussed above to the example
of the baker's map, which has been used as a prototype example for 
studying the semiclassics of chaotic maps in many investigations in
recent years \cite{Sar90,Ozo91,Sar92,Eck94,Sar94,Tan99}. 
The classical baker's map acts on points $(p,q)$ of the unit square
$[0,1]\times [0,1]$ according to
\begin{eqnarray}
 q' &=& 2q\ {\rm mod}\ 1 \\
 p' &=& (p+[2q])/2 \; ,
\end{eqnarray}
where $[x]$ denotes the integer part of $x$.
The periodic orbits of the map can be described by a complete binary 
symbolic code.
Each orbit is characterized by a symbol string 
$\{\epsilon_1,\dots,\epsilon_n\}$, where $\epsilon_i \in \{0,1\}$.
The action associated with a periodic orbit is given by \cite{Ozo91}
\begin{equation}
 S_\nu = {\nu\bar\nu\over 2^n-1} \ \ {\rm mod}\ 1
\end{equation}
with the integers $\nu$ and $\bar\nu$ defined by
\begin{equation}
\label{nu}
 \nu = \sum_{k=1}^n \epsilon_k\, 2^{k-1}\; , \qquad
 \bar\nu=\sum_{k=1}^n \epsilon_k\, 2^{n-k} \; .
\end{equation}
Each orbit of length $n$ has stability (i.e., largest eigenvalue of the 
monodromy matrix) $2^n$. 

A quantized version of the baker's map which preserves the classical 
symmetries was derived by Saraceno \cite{Sar90}. 
For even dimension $N$ of the Hilbert space the quantum map is given
by the unitary matrix
\begin{equation}
\label{defU}
U(N)=F_N^{-1}\times
\left(
\begin{array}{cc}
F_{N/2} & 0 \\
0 & F_{N/2}
\end{array}
\right)
\end{equation}
with
\begin{equation}
(F_N)_{nm}={1\over \sqrt N}\, 
\exp\biggl[-2\pi {\rm i}\Bigl(n-{1\over 2}\Bigr)\Bigl(m-{1\over 2}\Bigr)\biggr]
\qquad n,m=1,\dots,N\; .
\end{equation}
We will use the the exact quantum eigenvalues for comparison with the 
semiclassical ones obtained by harmonic inversion of the semiclassical signal
(\ref{Csemi}). 

\begin{figure}[p]
\vspace{19.5cm}
\includegraphics{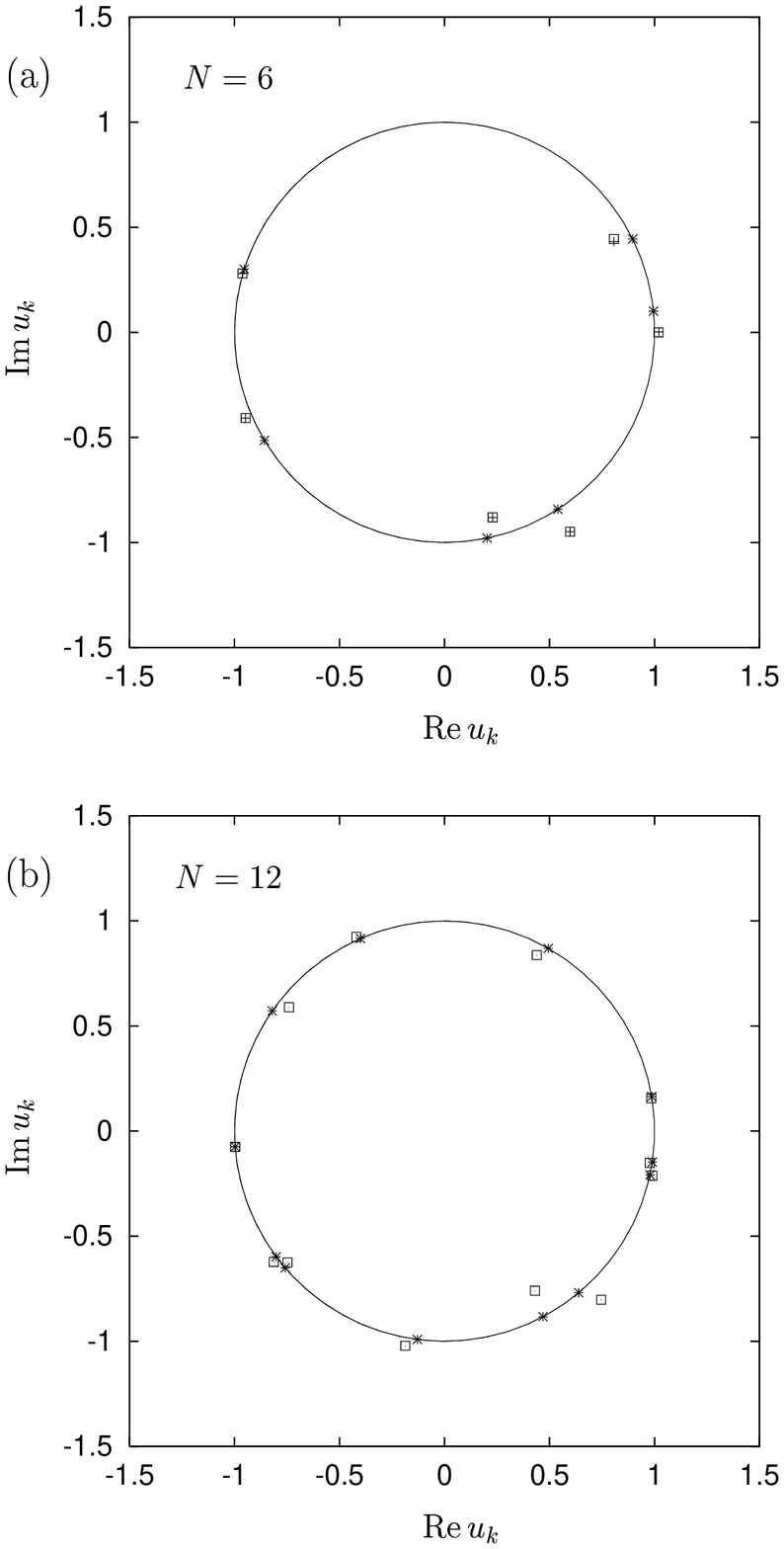}
\caption{Exact quantum ({\large $*$}) and semiclassical ($\boxdot$)
eigenvalues $u_k=\exp(-{\rm i}\varphi_k)$ of the baker's map obtained by 
harmonic inversion for dimensions (a) $N=6$ and (b) $N=12$. 
In the case $N=6$ the semiclassical results from 
Ref.~\cite{Eck94} are also plotted ($+$).}
\label{fig1}
\end{figure}

For the baker's map, the semiclassical amplitudes ${\mathcal A}_n$ in the 
signal (\ref{Csemi}) as defined in Eq.~(\ref{trU}) read \cite{Eck94}
\begin{equation}
 {\mathcal A}_n=-{\rm i}\sum_{{\rm po}(n)}
 {2^{n/2}\, n_0\over 2^n-1}\,\exp\left(
 2\pi {\rm i} N {\nu\bar\nu\over 2^n-1}\right)\; 
\end{equation}
with $\nu$ and $\bar\nu$ given by Eq.~(\ref{nu}).
We have performed calculations for dimensions $N=6$ and $N=12$.
The case $N=6$ has also been examined in Ref.~\cite{Eck94}, where the 
semiclassical eigenvalues were determined by a resummation of the 
Selberg zeta function of the map.
For the construction of the semiclassical signal, we calculated all orbits 
up to symbol length $n=20$ for dimension $N=6$ and up to length $n=38$ for
$N=12$.
Figure \ref{fig1} shows the results for the eigenvalues 
$u_k=\exp(-{\rm i}\varphi_k)$ obtained
by harmonic inversion of the semiclassical signal ($\boxdot$), compared with 
the exact quantum results ({\large $*$}) obtained by diagonalization of the 
quantum matrix $U$ (see Eq.~(\ref{defU})). 
For comparison, in the case $N=6$, we have also plotted the
semiclassical results from Ref.~\cite{Eck94} ($+$).

Our semiclassical results are in good agreement with the exact quantum
eigenvalues. 
However, as was also pointed out in Ref.~\cite{Eck94}, in the case of the
baker's map the semiclassical error is relatively large, which is probably 
due to the discontinuities inherent in this map.
In particular, a few of the semiclassical eigenvalues are located 
away from the unit circle at distances on the order of $10^{-1}$.
Although the accuracy of the semiclassical approximation should
improve in general with increasing $N$, the doubling of the dimension of
the Hilbert space from $N=6$ to $N=12$ on average does not yet visibly lead
the semiclassical eigenvalues closer to the unit circle.

On the other hand, in the case $N=6$, the semiclassical eigenvalues 
obtained by harmonic inversion are in excellent agreement with those 
from Ref.~\cite{Eck94}.
This implies that the deviations from the exact quantum eigenvalues
are indeed solely due to the semiclassical error, and do not indicate
any inaccuracies of the individual methods applied.
We add that the semiclassical approximation for the baker's map may even
be improved by introducing non-semiclassical correction factors to the 
amplitudes ${\mathcal A}_n$ \cite{Eck94,Sar94}.

As for flows, the harmonic inversion method for maps can also be used,
vice versa, to analyze the quantum spectrum in terms of the amplitudes
${\mathcal A}_n$ \cite{Mai97a}.
This is achieved by adjusting the quantum response function
(\ref{gqm}) to its semiclassical approximation (\ref{gsemi}).
However, since for maps the amplitudes ${\mathcal A}_n$ contain
contributions from all periodic orbits with topological length
$n$, it is not possible to extract information about single
periodic orbits from the quantum spectrum.
Moreover, since the traces ${\rm Tr}\,U^n$ do not depend on the quantity 
$\varphi$ which is to be quantized, the exact expression (\ref{gqm1}) for 
maps already fulfills the ansatz of the harmonic inversion procedure as 
a whole.
This is in contrast to scaling Hamiltonian systems, where the higher order 
$\hbar$ corrections to the semiclassical approximation depend on the scaling 
parameter (e.g., for billiard systems, the wave number) and the harmonic 
inversion of the exact quantum spectrum yields only the zeroth order $\hbar$ 
contributions to the response function, with the higher orders acting as a 
kind of noise \cite{Mai99a,Wei00}.
As a consequence, the analysis of the quantum spectrum of maps
will simply yield the exact quantum values for the traces ${\rm Tr}\,U^n$ 
rather than their semiclassical approximations ${\mathcal A}_n$
(we have verified this for the baker's map), and no information about the 
semiclassics or single periodic orbits can be obtained.

In conclusion, we have presented a method for the periodic orbit quantization
of chaotic maps, which makes use of harmonic inversion.
The procedure works similar to the one for flows, and can in the same way 
be applied to all chaotic maps, independent of any special properties of 
the respective system.
We have demonstrated the power of the method by successfully applying it to
the baker's map.
Our results reproduce the exact quantum eigenvalues of the baker's map
to within the error of the semiclassical approximation, and are in 
excellent agreement with those obtained by other semiclassical methods.

We gratefully acknowledge support of this work by the Deutsche 
For\-schungs\-ge\-mein\-schaft.


\end{document}